\documentclass[%
reprint,
amsmath,amssymb,
prb,
]{revtex4-2}

\usepackage{graphicx}
\usepackage{dcolumn}
\usepackage{bm}
\usepackage{subfigure}
\usepackage{booktabs}
\usepackage[colorlinks=true, letterpaper=ture, pdfstartview=FitV, linkcolor=blue, citecolor=blue, urlcolor=blue]{hyperref}

\begin{document}

\preprint{APS/123-QED}

\title{Atomistic simulations of thermodynamic properties of liquid gallium from first principles}

\author{Hongyu Wu$^{1,2}$}
\author{Wenliang Shi$^{1,2}$}
\author{Ri He$^{1,2}$}
\author{Guoyong Shi$^{3}$}
\author{Chunxiao Zhang$^{4}$}
\author{Zhicheng Zhong$^{1,2}$}
\thanks{zhong@nimte.ac.cn}
\author{Run-Wei Li$^{1,2}$}

\affiliation{\\$^1$CAS Key Laboratory of Magnetic Materials and Devices, and Zhejiang Province Key Laboratory of Magnetic Materials and Application Technology, Ningbo Institute of Materials Technology and Engineering, Chinese Academy of Sciences, Ningbo 315201, China}

\affiliation{$^2$College of Materials Science and Opto-Electronic Technology, University of Chinese Academy of Sciences, Beijing 100049, China}

\affiliation{$^3$Department of Physics, Yantai University, Yantai, 264005, China}

\affiliation{$^4$School of Physics and Optoelectronic Engineering, Shandong University of Technology, Zibo, Shandong 255100, China}


\date{\today}

\begin{abstract}
Thermodynamic properties serve as essential characteristics of any physical systems, offering insights into the system's behavior under specific conditions. Determining these properties in disordered systems from first principles remains a formidable challenge due to the intricate incorporation of nuclear quantum effects into large-scale atomic simulations.
In this study, we employ a machine-learning deep potential model in conjunction with the quantum thermal bath method to address the complexities of nuclear quantum effects. Applying this approach, we accurately calculate the variations in various thermodynamic quantities of liquid gallium across temperatures ranging from zero to room temperature. These quantities include internal energy, specific heat, enthalpy change, entropy, and Gibbs free energy, demonstrating excellent agreement with experimental data.
Our research marks a significant advancement in the exploration of thermodynamics within liquids, amorphous materials, and other disordered systems.
\end{abstract}

\maketitle


\section{Introduction}
Compared to crystalline structures, simulating the complex and disordered structures of liquids, particularly their thermodynamic properties, has posed a long-standing and formidable challenge\cite{Massobrio2015}. Conventional simulation methods based on classical statistical only yield results consistent with the equipartition theorem when describing thermodynamic properties\cite{Plimpton1995}. According to the equipartition theorem, the specific heat of a solid should be a constant, which contradicts the Debye-Einstein solid specific heat theory\cite{Kittel1996}. In fact, the equipartition theorem does not strictly apply to the energy distribution in liquids either\cite{Greywall1983}. In order to accurately compute quantities such as internal energy and specific heat, particularly at low temperatures, it is necessary to incorporate quantum statistics into simulations. However, due to the absence of ordered atomic arrangements in disordered systems, theories relying on harmonic approximation in crystals will no longer be applicable\cite{Kittel1996}. As a result, the challenge primarily stems from the inherent difficulty in incorporating nuclear quantum effects through phonon theory\cite{Ceperley1995}, it fundamentally complicates the accurate calculation of thermodynamic properties. For example, one widely accepted method to incorporate nuclear quantum effects is path integral molecular dynamincs, it suffers from extremely expensive calculation cost\cite{Marx1994,Marx1996}. When it comes to statistics related issues, results of this method are difficult to achieve statistical equilibrium due to scale constraints. Therefore, a universal and computationally efficient approach for introducing nuclear quantum effects that can be applied in large-scale simulations of liquids is highly necessary. 

\begin{figure*}[htbp]
\centering
\subfigure{
\includegraphics[width=\columnwidth]{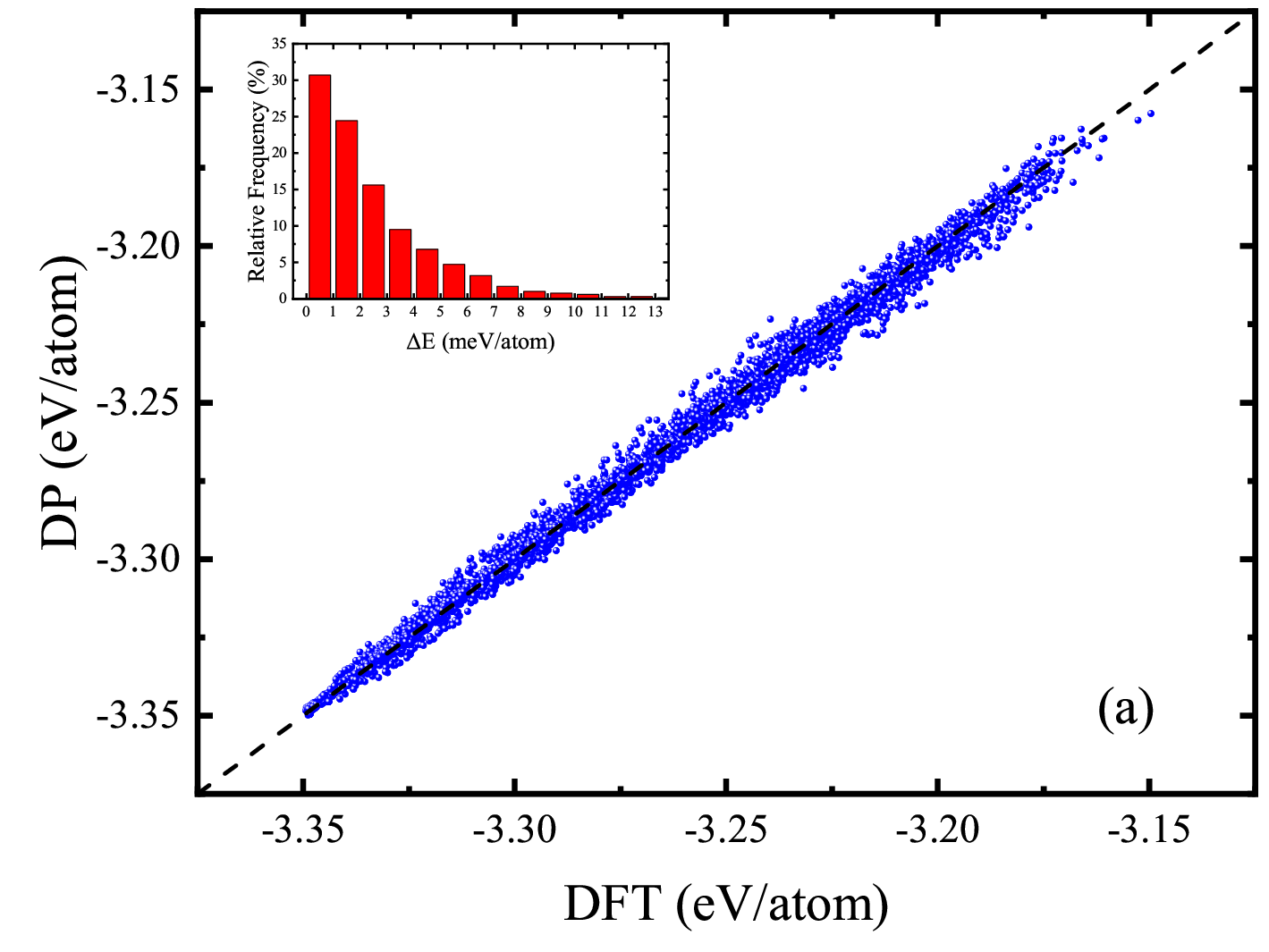}
\label{1a}
}
\subfigure{
\includegraphics[width=\columnwidth]{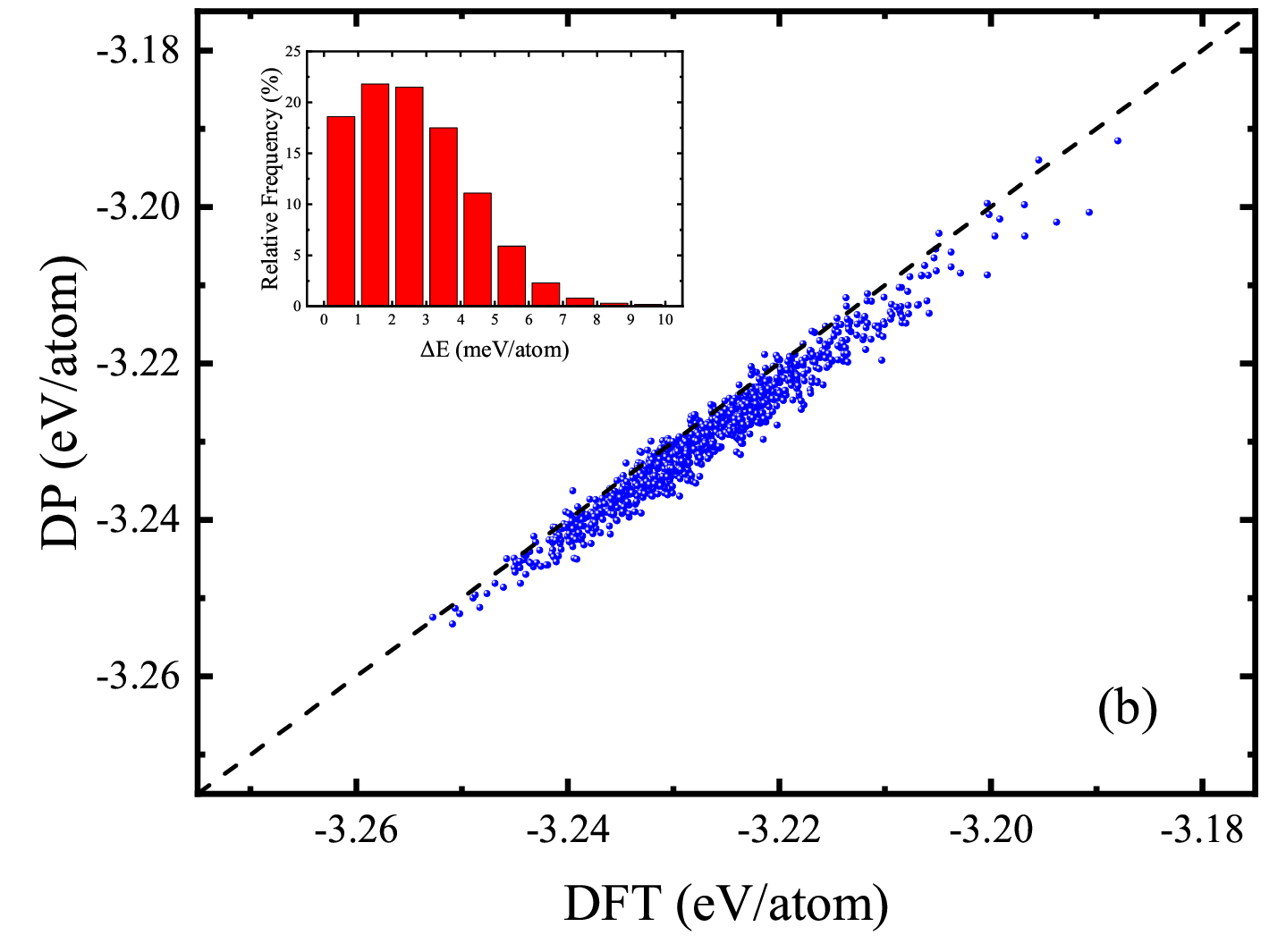}
\label{1b}
}
\quad
\subfigure{
\includegraphics[width=\columnwidth]{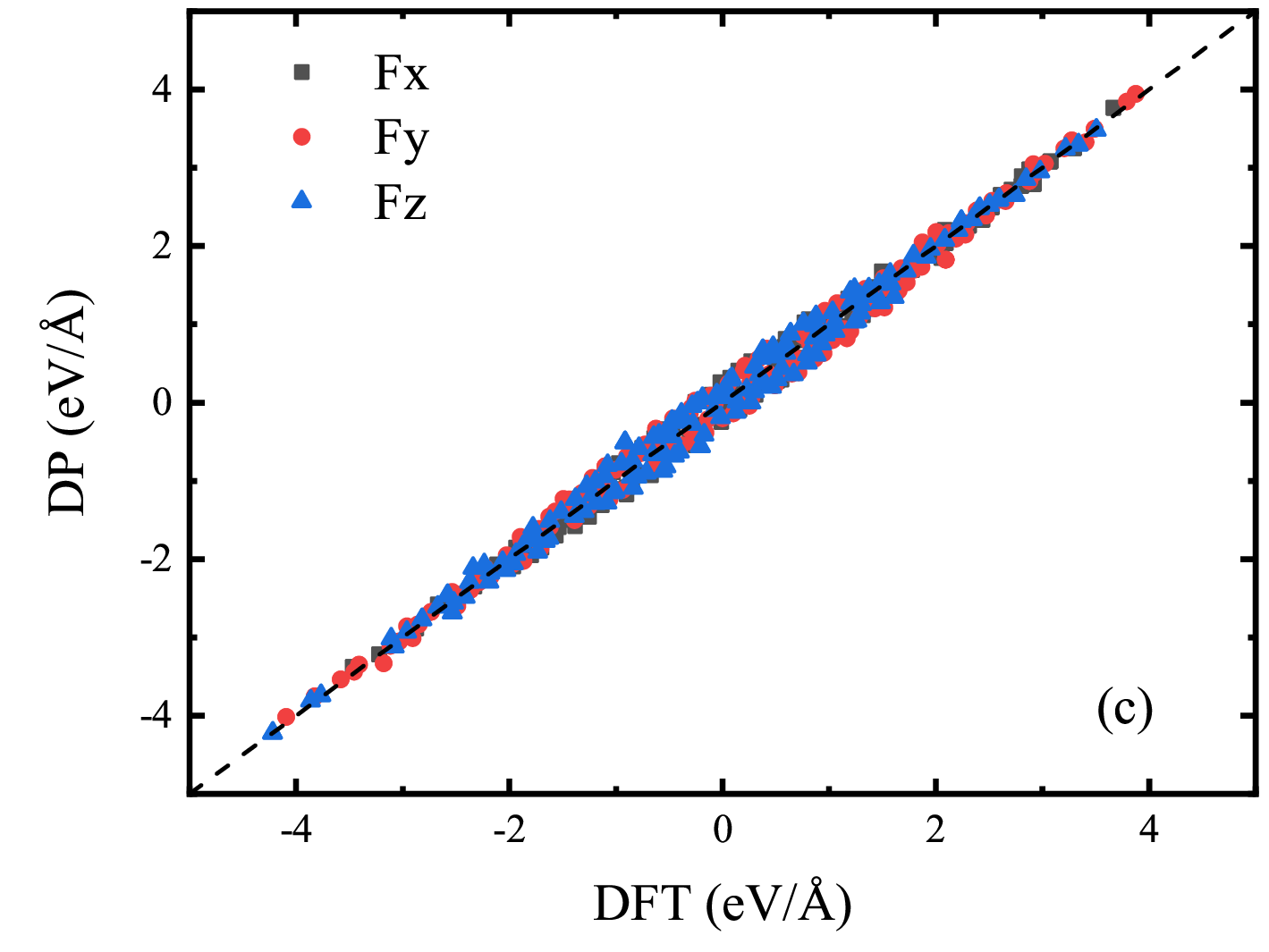}
\label{1c}
}
\subfigure{
\includegraphics[width=\columnwidth]{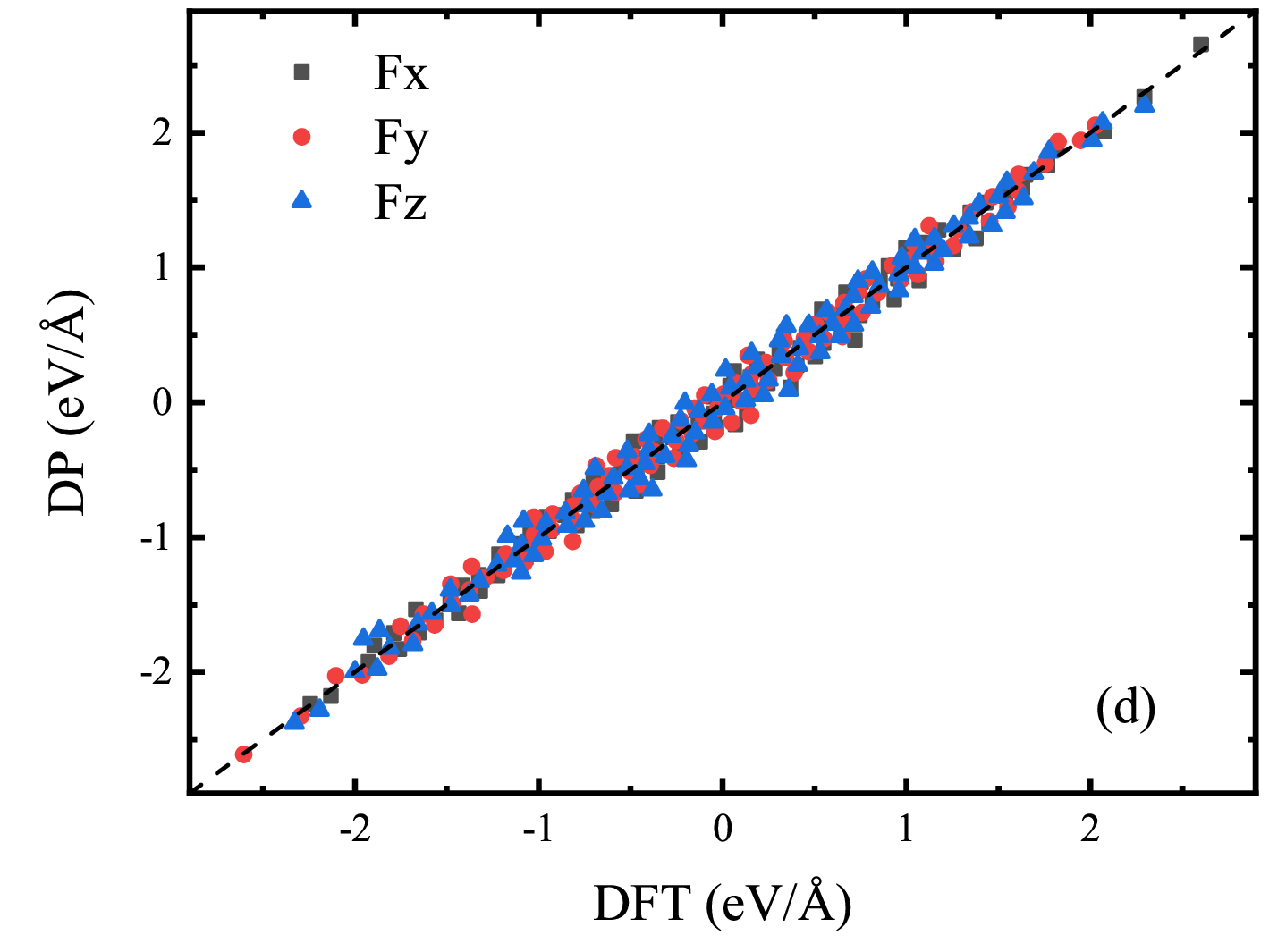}
\label{1d}
}
\caption{Comparison of (a) and (b) energies (c) and (d) atomic forces using the DFT and DP model, (a) and (c) are the final training dataset (including order and disorder configurations), (b) and (d) are the testing datasets of disorder configurations, respectively.}
\label{modeltest}
\end{figure*} 

One other challenge for classical simulations of disorder system is the precision of force fields\cite{Miller2005}, these force fields are often fitted to experiments and the choice of them can affect simulation results\cite{Becker2013}. Meanwhile, the scale and precision in simulating disorder systems both affect the accuracy of the results\cite{Gunsteren1998,Zhang2021,Sosso2016,Niu2020}. Therefore, a prerequisite for introducing nuclear quantum effects in simulating disordered liquid systems is ensuring the accuracy of the description of atomic interactions.

To overcome these two limits in simulation of disorder systems, we use the deep potential (DP) + quantum thermal bath (QTB) approach which enables large-scale atomistic dynamic simulations with the precision of Density Functional Theory (DFT). This approach is both \textit{ab initio} and efficient, and it has been demonstrated to accurately capture the complex phenomena arising from nuclear quantum effects in crystals in our previous works\cite{Wu2022,Qin2023}. DP is a machine-learning method that generates accurate force fields for Molecular Dynamics (MD) by sampling DFT results\cite{Zhang2018,Zhang2018a}. QTB is a method that preserves the quantum statistical features in MD simulations\cite{Dammak2009}. Its fundamental concept involves transforming the basic Newtonian equations in MD into Langevin-like equations, and incorporating colored noise modified with quantum fluctuation and dissipation theorem\cite{Callen1951}. Since QTB corrects the total forces in MD, it can be implemented even without periodic boundary conditions, enabling the introduction of nuclear quantum effects into disordered liquid systems. The DP+QTB approach requires a computational cost comparable to classical MD, making it capable of simulating systems with millions of atoms, yielding sufficient physical information.

To demonstrate the power of DP+QTB in situation of simulating disorder systems, we focus on the gallium (Ga). Ga is an element that exhibits a number of unique and untrivial properties that make it an attractive material for scientific research. Among its characteristics, its low melting temperature ($\sim$ 300 K)\cite{Sostman1977} make it a promising candidate for studying solid-liquid transition and thermodynamics in liquids. Ga and its alloys have recieved attentions in many field of applications, such as soft electronics, energy harvesting, etc\cite{Lin2020,Song2020,Tang2021}. Due to the extremely low melting point of Ga, the influence of nuclear quantum effects on thermodynamic properties near its melting point is more pronounced compared to other metals and cannot be ignored.

In this study, we investigate the dynamics and thermodynamic properties of Ga using the DP+QTB approach. Various dynamic phenomena of Ga, such as the phase transition and atomic diffusion, their DP+QTB results are all comparable with experimental measurements. Furthermore, it's worth to note that since the DP+QTB approach is based on first principles and incorporates nuclear quantum effects, we can exactly compute various thermodynamic quantities for both solid and liquid Ga. These quantities include internal energy, specific heat, entropy, enthalpy change and Gibbs free energy. Based on the results obtained, we validate that the DP+QTB is a feasible approach for investigating the thermodynamics and dynamics of disordered systems, including liquids, amorphous materials, and more.

\section{computational methods}

\subsection{Deep Potential of Ga}
DP is a machine-learning-based method that aims to provide highly applicable and accurate interatomic interaction potentials\cite{Zhang2018a}. The core concept of the DP model is to fit the energy and force of different configurations of a specific material using a deep neural network. The training data consists of numerous structural configurations (typically containing up to 50 atoms) and corresponding high-precision DFT results for energy and force. Once the model is well-trained, it can accurately predict the total energy and atomic forces for any given configuration, even those not present in the training data or involving large supercells, all while minimizing computational costs. This DFT-level accuracy makes the DP model suitable for application as a force field in MD simulations. The DP method has successful applications in the metal system such as Cu\cite{Zhang2020}, Al-Mg binary alloys\cite{Zhang2019}, etc.

To construct the DP model specific to Ga, we carefully curated a training dataset consisting of 4227 candidate configurations. These configurations encompassed various phases of Ga, including $\alpha$, $\beta$, and II phases, and they include both ordered and disordered configurations. The generation of this dataset involved performing MD simulations in the isobaric-isothermal (NPT) ensemble, with temperatures ranging from 0 to 600 K and pressures ranging from 0.0001 to 5 GPa. For a more comprehensive understanding of the DP methodology and the details of dataset generation, please see the supplementary materials\cite{Supplementary}.

The comparison between the DP model predictions and DFT results is illustrated in FIG.~\ref{modeltest}, which showcases the error analysis. The final training dataset yield energy Root Mean Square Error (RMSE) of 3.37 meV/atom and force RMSE of 0.049 eV/\AA, demonstrating the accuracy of the DP model, as shown in FIG.~\ref{1a}\&\ref{1c}. To further assess the DP model's capability in predicting liquid Ga, we conducted a comparison using disorder configurations not included in the final training dataset. As shown in FIG.~\ref{1b}\&\ref{1d}, the results exhibit consistency between DFT and the DP model, with RMSE values of 3.15 meV/atom for energy and 0.049 eV/\AA for force, respectively. 

\subsection{Quantum Thermal Bath}
DP potential is able to produce DFT-level energy and atomic force in MD simulation, and we use QTB to incorporated quantum effects in MD simulations. In MD classical limit, the equipartition theorem is fulfilled, therefore it only produce results that confirm to classical behavior\cite{Plimpton1995}. The core idea of QTB is based on the quantum mechanical fluctuation dissipation theorem\cite{Callen1951}, it introduces associated random force and friction term into the equation, constituting a quantum thermal bath\cite{Dammak2009}. The equation of motion of a degree of freedom $x$ of a particle of mass $m$ in the presence of an external DP force $F(x)$ is modified to Langevin-like equation
\begin{equation}
m\ddot{x}=F(x)+\sqrt{2m\gamma}\Theta(t)-\gamma m\dot{x},
\end{equation}
where $\Theta(t)$ is a colored noise with a power spectral density
\begin{equation}
\Theta(\omega,T)=\hbar\omega[\frac{1}{2}+\frac{1}{exp(\frac{\hbar\omega}{k_BT})-1}],
\label{PSD}
\end{equation}
it includes the zero-point energy. 

QTB can be easily manipulated and is independent of the studied system\cite{Barrat2011}. Previous works have demonstrated that QTB can produce results in good agreement with experiments or widely accepted but highly resource-intensive simulation methods results for single crystal systems\cite{Wu2022,Dammak2009,Barrat2011,Dammak2011,Qin2023}, and computation complexity comparable to classical MD. DP+QTB can provide exact physical quantities for harmonic systems such as single crystal. In cases involving anharmonic situations like liquids and amorphous solids, simulations using the QTB method may tend to exhibit some errors\cite{Dammak2011}.

\section{results and discussions}
\subsection{Dynamics properties}

\begin{figure}[t]
\centering
\includegraphics[width=\columnwidth]{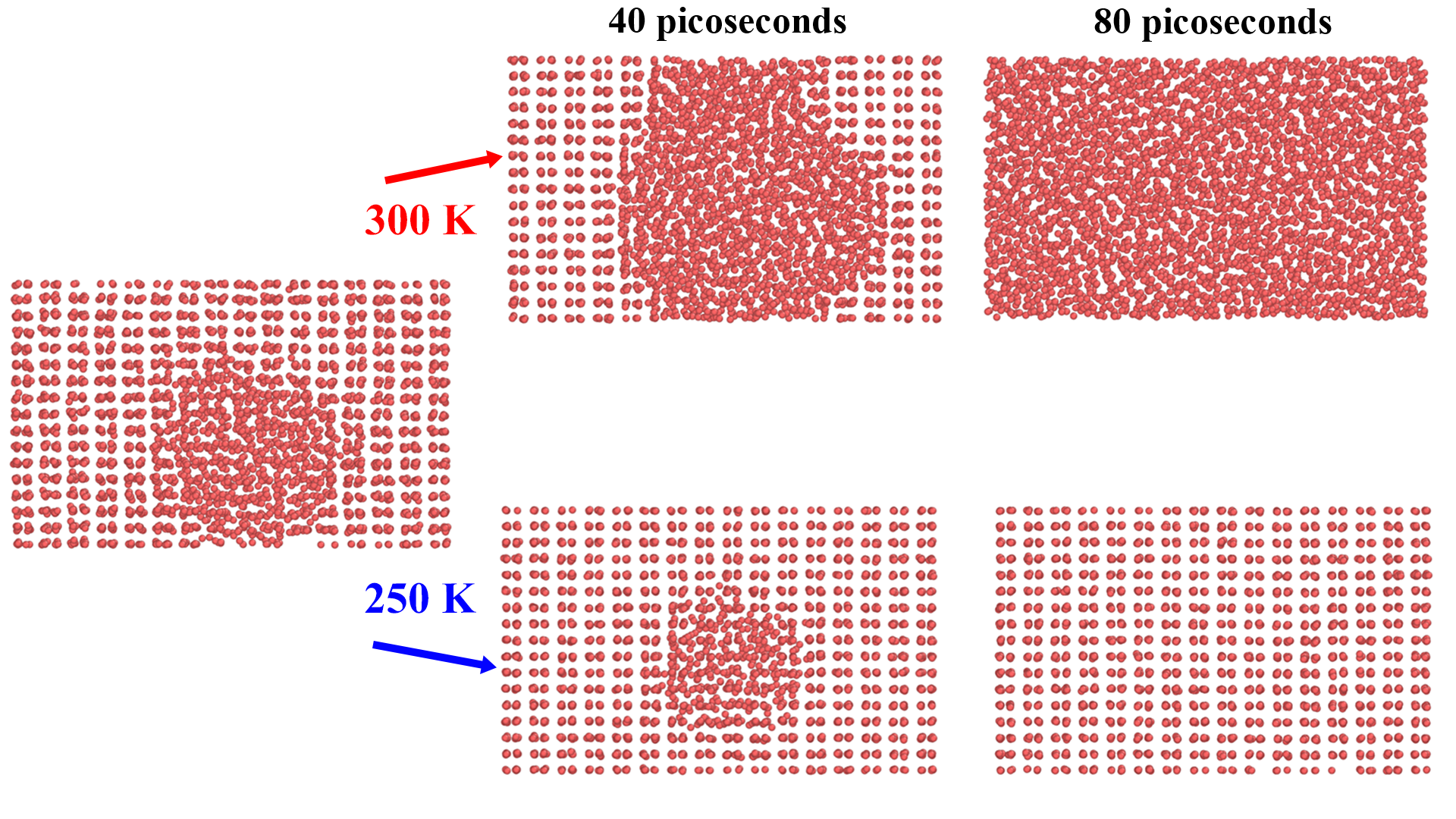}
\caption{Schematic of melting core and its heating and cooling processes, respectively.}
\label{Core}
\end{figure}

One of the most important and practical properties of Ga is its relatively low melting point, allowing it to melt into a liquid state at room temperature. The primary challenge of our DP model lies in accurately describing the solid-liquid transition of Ga. Unlike the experiment, the advantage of MD is it can track the motion of each atom, making it easy to study the dynamics of atoms in condensed matters. For instance, we can directly observe each atoms and its structural changes. As shown in FIG.~\ref{Core}, considering the nucleation dynamics, we introduced the concept of a melting nucleus. The melting nucleus is a local disorder structure. The reason for introducing the concept of melting nuclei is that, in practical simulation, it is challenging to directly simulate phase transitions near the melting point. Especially during the solidification process, without the introduction of nucleation, the entire structure would remain in a supercooled liquid state. Subsequently, in this work, the solids at temperatures higher than the phase transition point and the liquids at temperatures lower than the phase transition point are referred to as superheated solids and supercooled liquids, respectively. This nucleus exhibited growth at 300 K and shrinkage at 250 K, corresponding to the processes of melting and crystallization, respectively. Further, we utilize a numerical approach to illustrate the solid-liquid phase transition point of Ga in FIG.S2\cite{Supplementary}. The exact melting point obtained by DP is 292 K, and after incorporating quantum effect is 270 K, and they are close to the actual melting point of Ga\cite{Sostman1977}. The melting point of Ga calculated using DP+QTB is slightly reduced, which is attributed to the contribution of zero-point energy, resulting in a slightly lower required phase transition temperature. The lattice constants of $\alpha$ phase Ga at 300K calculated by DP are listed in Table~\ref{EXP}, and they are comparable to the experimental results\cite{Barrett1965}.

\begin{table}[b]
\centering
\caption{Phase transition point($T_c$/K); lattice constants(a,b,c/\AA), self diffusion coeffcient (D/$10^{-5}cm^2/sec$, FIG.~\ref{MSDSCL}), expansition of solidification ($\Delta V$, \%) and heating enthalpy change ($\Delta H$/Jg$^{-1}$, FIG.~\ref{Etotal}) at 300 K; specific heat/3Nk$_B$ before and after melting ($C_b$ \& $C_a$/Jkg$^{-1}$K$^{-1}$, FIG.~\ref{Cboth}) of Ga by experiments and DP+QTB.}
\label{EXP}
\begin{ruledtabular}
\begin{tabular}{ccc}
\textrm{}  & \textrm{Exp.}          & \textrm{DP+QTB} \\
\colrule
$T_c$ &  303\cite{Sostman1977}  & 270\\
a & 4.5258\cite{Barrett1965}                              & 4.5075 \\
b & 4.518\cite{Barrett1965}                               & 4.4965 \\
c & 7.6602\cite{Barrett1965}                              & 7.6384 \\
$\Delta V$ & 3.1 & 3.6 \\
$D$ & 1.70\cite{Petit1956}  & 2.69 \\
$\Delta H$ & -74\cite{Kumar2015}, -80\cite{He2005}  & -49\\
$C_b$ &1.077\cite{AdamsJr1952}& 1.030\\
$C_a$ &1.166\cite{AdamsJr1952}& 1.116\\
\end{tabular}%
\end{ruledtabular}
\end{table}

\begin{figure}[htbp]
\centering
\subfigure{
	\includegraphics[width=\columnwidth]{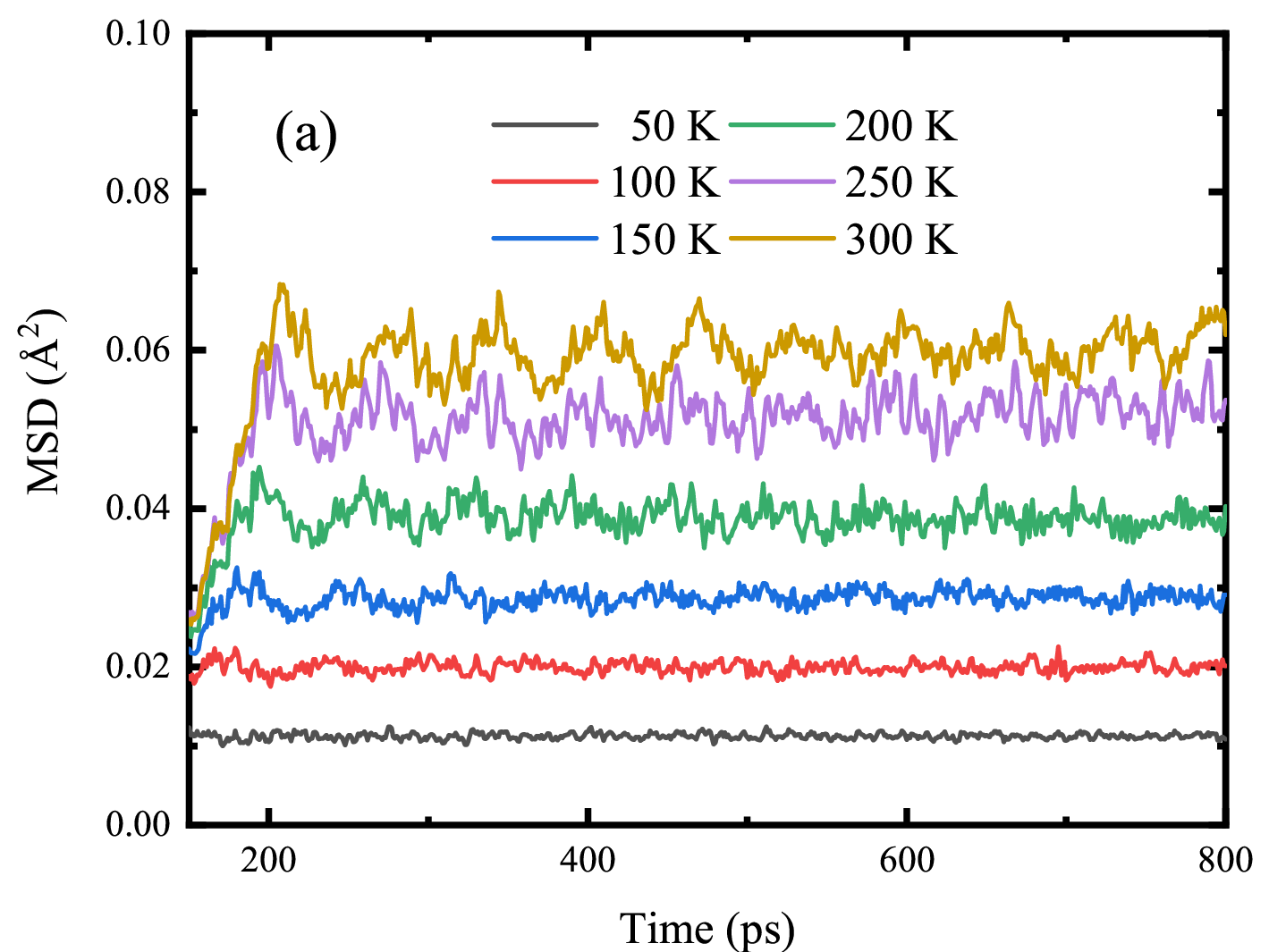}
	\label{MSDsolid}
}
\quad
\subfigure{
	\includegraphics[width=\columnwidth]{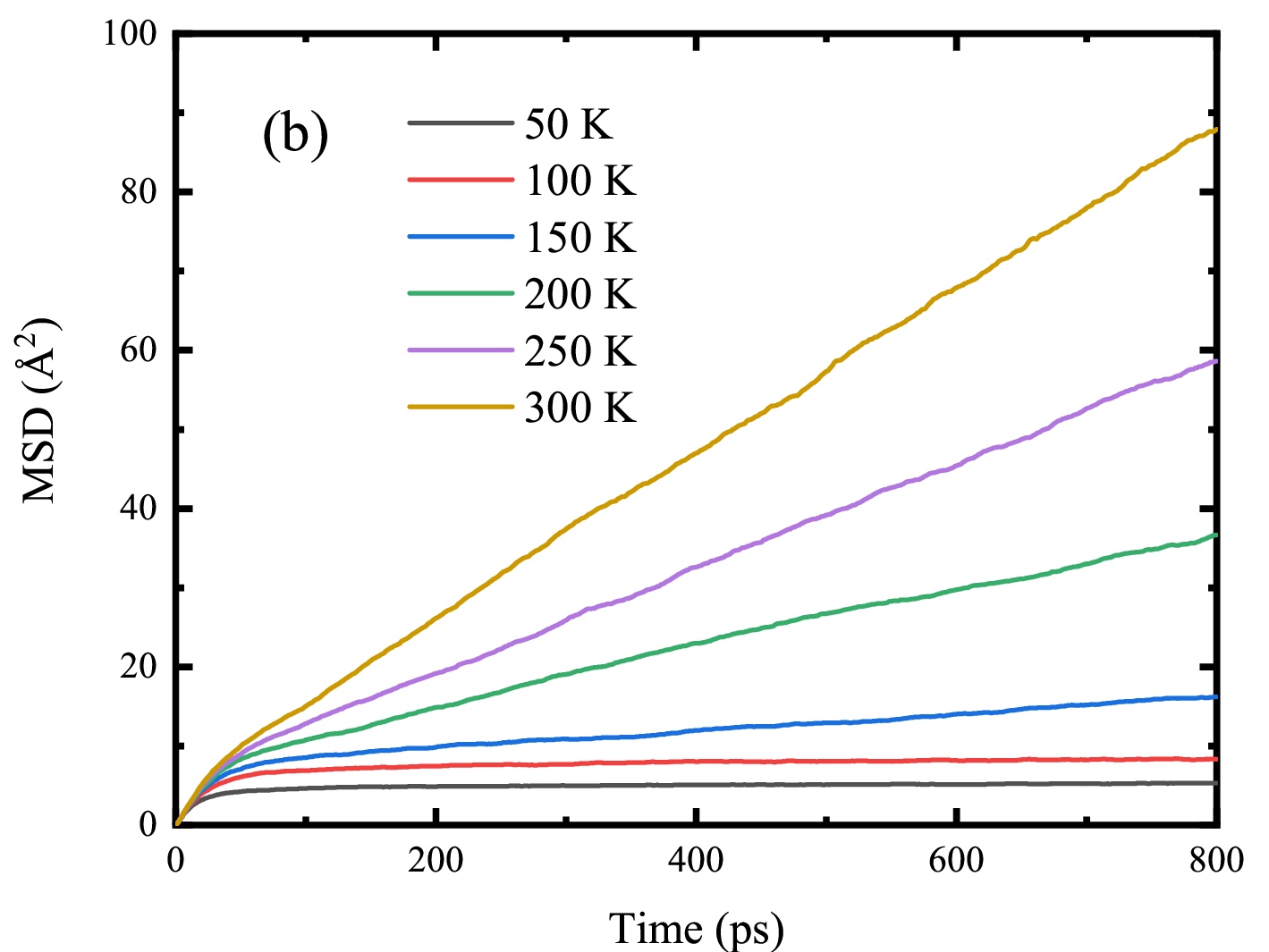}
	\label{MSDSCL}
}
\caption{Mean square displacement of (a) solid and (b) liquid Ga.}
\label{MSD}
\end{figure}

\begin{figure*}[htbp]
\centering
\subfigure{
\includegraphics[width=\columnwidth]{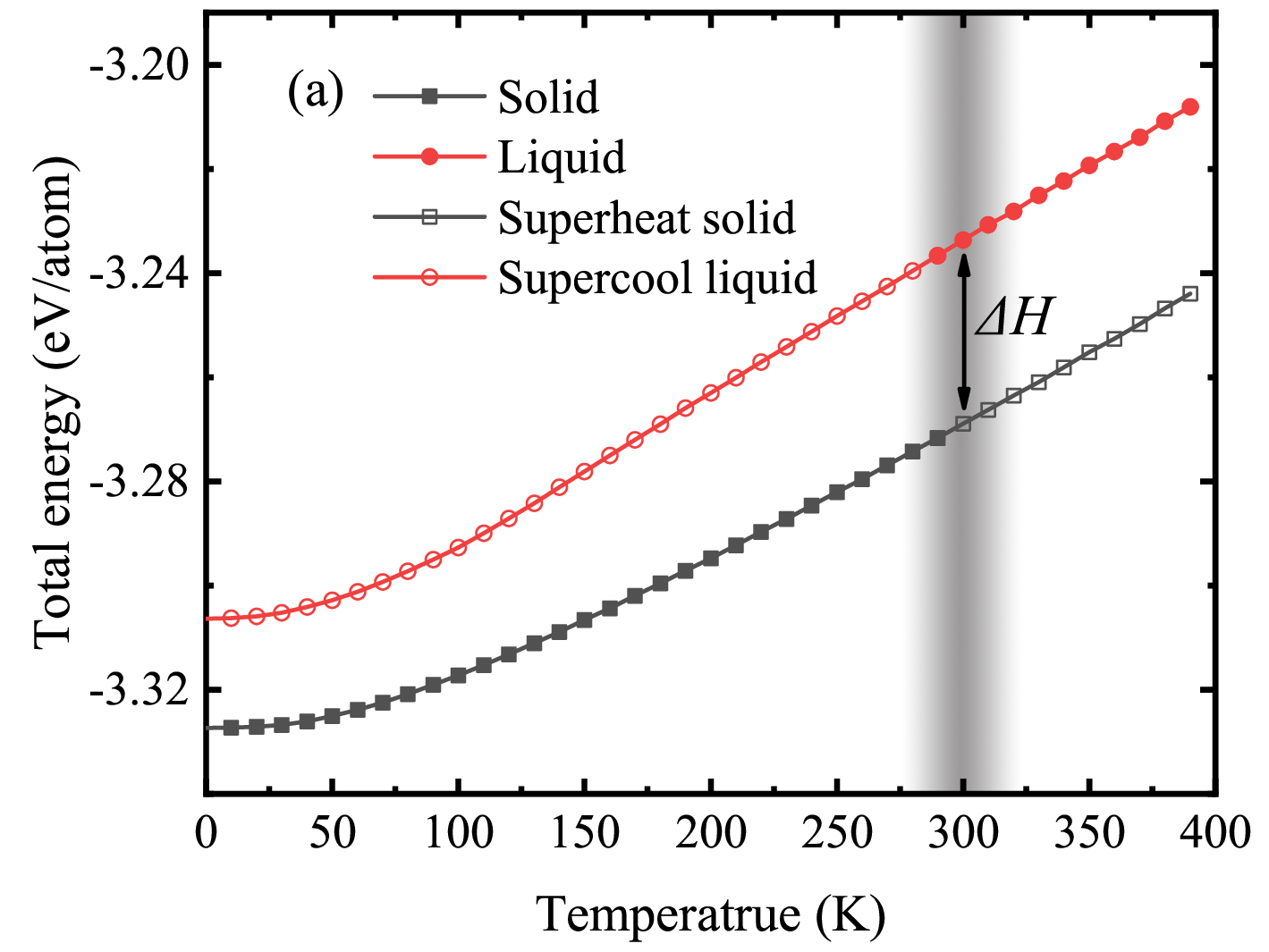}
\label{Etotal}
}
\subfigure{
\includegraphics[width=\columnwidth]{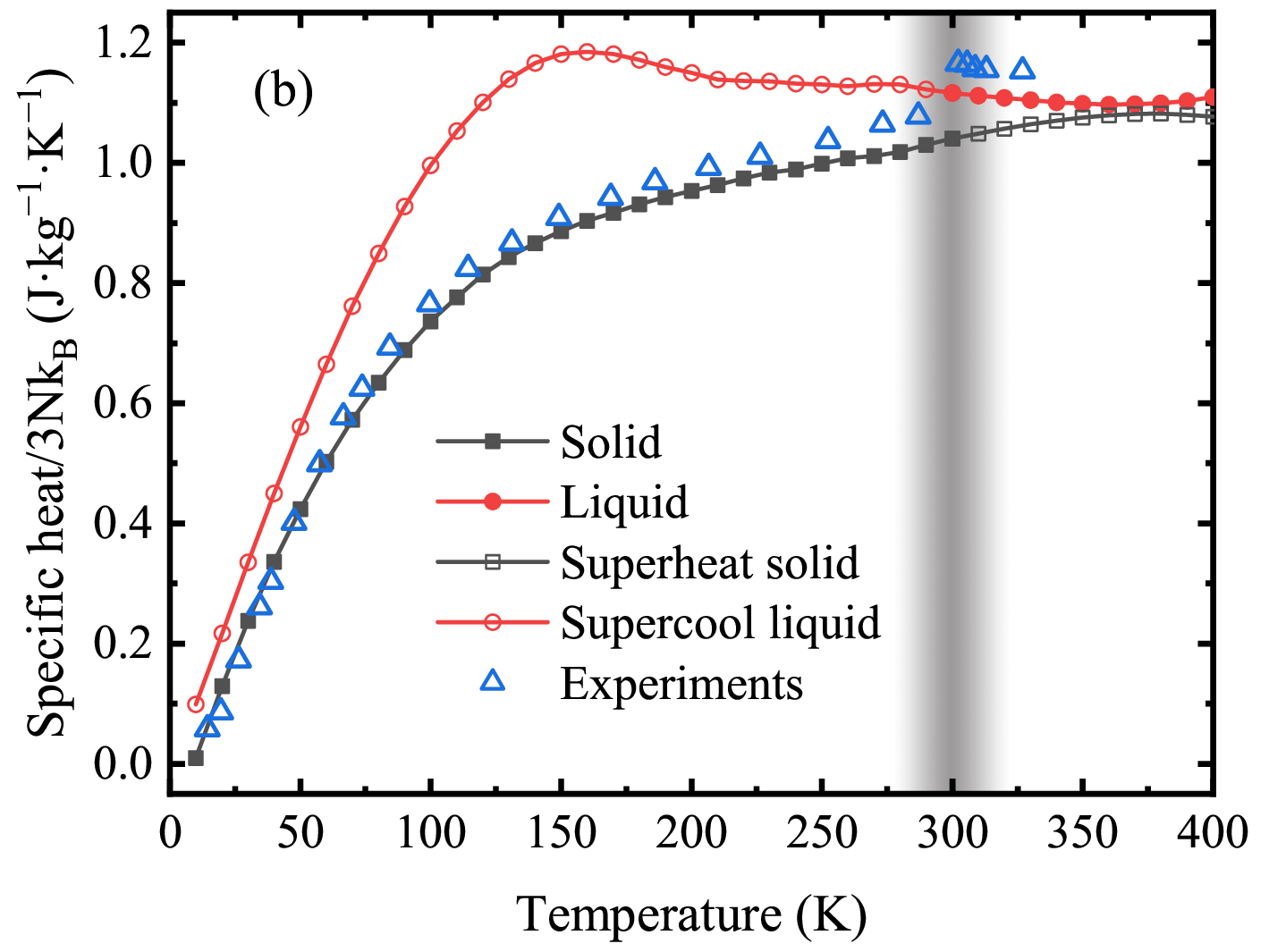}
\label{Cboth}
}
\caption{(a) Total energy and (b) specific heat at constant pressure in quantum condition of solid and liquid Ga as a function of temperature.}
\end{figure*}

The collective thermal motion of Ga atoms can be studied by examining the mean squared displacement (MSD):
\begin{equation}\left\langle \Delta r^2(t) \right\rangle = \frac{1}{N} \left\langle \sum_{j = 1}^N \left\lbrack \mathbf{r}_ j(t) - \mathbf{r}_ j(0) \right\rbrack^2 \right\rangle.
\end{equation} 
For solids, their structures remain crystalline, with each atom vibrating near its equilibrium position. Aside from small thermal fluctuations at high temperatures, their MSD remains a small constant value and does not vary with simulation time, as shown in FIG.~\ref{MSDsolid}. Compared to solids, the magnitude of MSD in liquids is significantly higher, indicating that atoms in liquids exhibit much greater mobility. And the MSD in liquids increases linearly with time, allowing us to perform a linear fit to obtain the self-diffusion coefficient $D=d\left\langle \Delta r^2(t) \right\rangle/dt$ of Ga. At 300 K, the $D$ under classical and quantum situations are 1.03 and 2.69 $10^{-5}cm^2/sec$, they have the same magnitude with experiments result (Table~\ref{EXP})\cite{Petit1956}. The results of the phase transition point and self diffusion coefficient indicate that our DP model exhibits good accuracy, as it can describe disordered liquid structure of Ga. For more details on the atomistic structure of liquid Ga described by our DP model, please refer to the section 3 of supplementary materials\cite{Supplementary}.

\subsection{Thermodynamic Properties}

In condensed matter systems, quantum effects become increasingly significant as temperatures decrease, making it impossible to neglect their influence on macroscopic physics. For example, the zero-point energy, which arises from the uncertainty principle in quantum mechanics, imposes a lower limit on the energy of condensed matter systems at absolute 0 K. 
The strength of the zero-point energy can be quantified by the definition of zero-point temperature\cite{Wu2022}, given by 
\begin{equation}
T_{zero}=\frac{1}{k_B}\int_{0}^{\infty}g(\omega)\frac{1}{2}\hbar\omega d\omega,
\label{Tzero}
\end{equation}
where $g(\omega)$ is the phonon density of states of Ga. This parameter signifies the average energy per atom associated with zero-point fluctuations represented in terms of temperature. The $T_{zero}$ of Ga is approximately around 100K, indicating a significant nuclear quantum effect. Therefore, this effect should not be neglected in our simulations.

The total energy calculated by DP+QTB of solid and liquid are shown in FIG.~\ref{Etotal}.
At finite temperatures, the increase in energy with temperature is not as significant in the quantum case as $3Nk_BT$ in the classical case. This behavior can be attributed to the energy level gap of quantum excitations, wherein only certain phonons contribute to the excitation process. At higher temperatures, the change in energy with temperature returns to a classical behavior. The energies of relaxed liquid structures are compared with solid, and their energies are higher than solid states about 30 meV/atom, and we show it in FIG.S7 in supplementary materials\cite{Supplementary}.
As shown in the black arrow line in FIG.~\ref{Etotal}, the energy difference between two states is the enthalpy change ($\Delta H$) between two states. The $\Delta H$ obtained by DP+QTB is -49 Jg$^{-1}$, and its magnitude is comparable to experimental results (Table.~\ref{EXP})\cite{Kumar2015,He2005}.

\begin{figure*}[htbp]
\centering
\subfigure{
\includegraphics[width=\columnwidth]{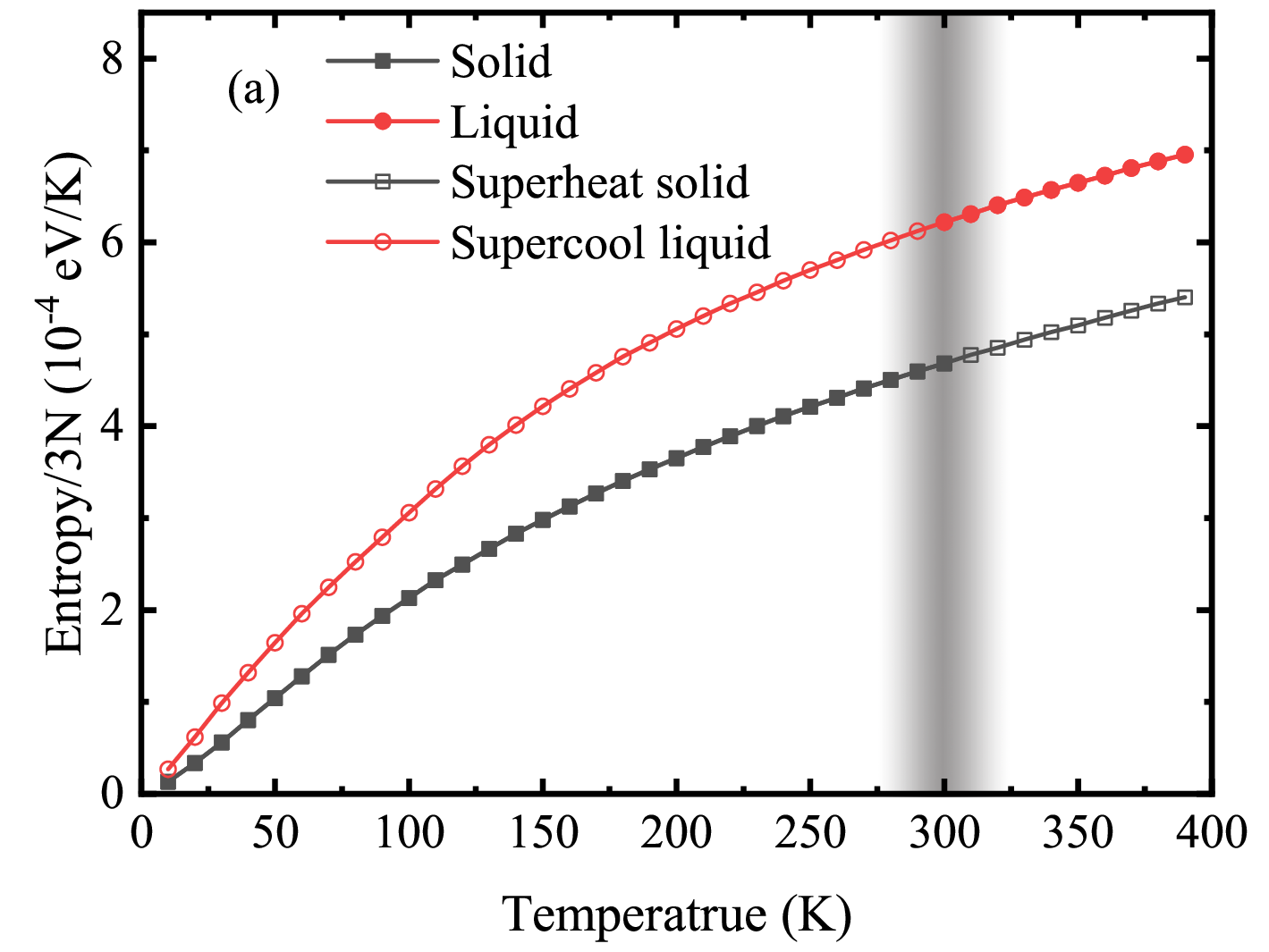}
\label{Entropy}
}
\subfigure{
\includegraphics[width=\columnwidth]{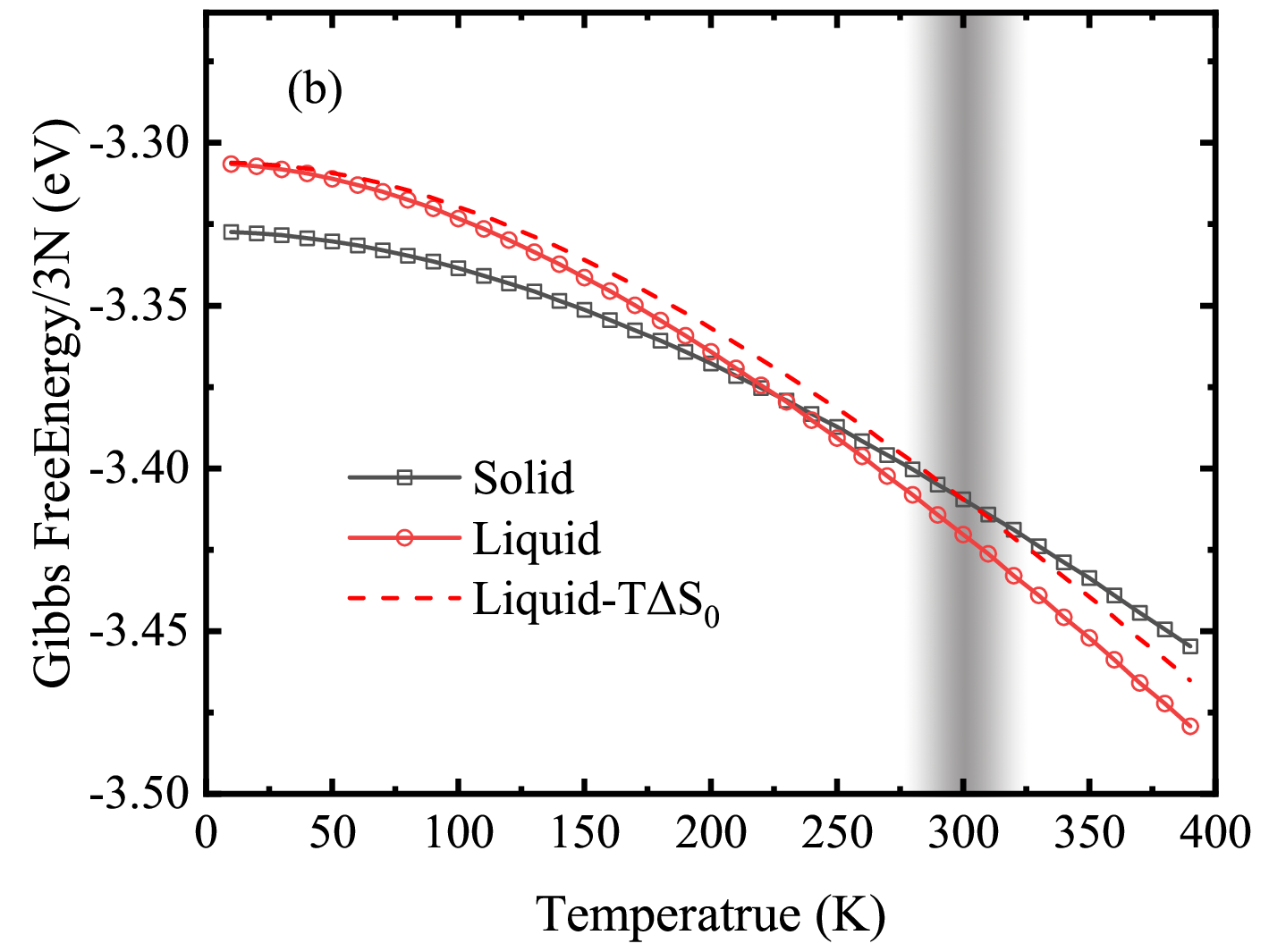}
\label{FreeEnergy}
}
\caption{Comparison of (a) entropy and (b) free energy of solid and liquid Ga}
\end{figure*}

After accurately computing the temperature dependence of energy $U(T)$ by considering nuclear quantum effects, the specific heat can be obtained by taking the first derivative of energy with respect to temperature $\partial U/\partial T$, as shown in black dots line in FIG.~\ref{Cboth}. We compared our results with experimental measurements\cite{AdamsJr1952}, as shown by the blue dot in FIG.~\ref{Cboth}, and found that the DP+QTB simulation results are in agreement with the experimental values, even at extremely low temperatures. Additionally, we have also calculated the specific heat of the liquid phase, as shown by the red dots line in the FIG.~\ref{Cboth}. Before the melting point, the specific heat of the liquid phase is consistently higher than that of the solid phase. This indicates that liquid exhibits more intense heat absorption and release, and the change in potential energy with temperature is also more significant. When the temperature exceeds the melting point, the specific heat of the liquid aligns with the experimentally measurements in both magnitude and decreasing trend. Furthermore, as the temperature gradually increases, the specific heats of the solid and liquid gradually approach each other. Furthermore, it can be observed from FIG.\ref{Cboth} that there are still noticeable deviations between our simulations and experimental datas, especially as temperature increases, this discrepancy becomes more pronounced. This is likely due to our omission of the electronic specific heat in our simulations. Electronic specific heat increases linearly with temperature\cite{Kittel1996}, and as a result, this error becomes more prominent as the temperature rises.

As for the reason for the higher specific heat of the liquid, we believe it's because the liquid phase has fewer high-frequency phonon modes around the equilibrium positions compared to the solid phase, while it has more low-frequency phonon modes with larger diffusion distances. These low-frequency modes are easier to excite, which leads to the liquid exhibiting stronger heat-absorbing capacity. To validate this hypothesis, we conducted calculations on the phonon spectrum of a super-large disordered Ga liquid structure, as shown in FIG.S8 in the supplementary materials\cite{Supplementary}. Compared to the calculation for the crystal, it is indeed evident that the high-frequency modes disappear, and the low-frequency modes increase, thereby confirming our hypothesis.

The Gibbs free energy is a crucial thermodynamic quantity used to assess the stability of phases. However, calculating the Gibbs free energy involves determining entropy, which is notoriously difficult to compute accurately. According to statistical physics, calculating entropy requires considering all possible microscopic states. Since we can accurately calculate the internal energy $U$ and the constant-pressure heat capacity $C_P(T)$ using the DP+QTB method, we will use thermodynamic equations to compute entropy and, consequently, the Gibbs free energy.
Since the initial entropy needs to be rigorously determined by traversing all configurations in the phase space, we can only assume that entropy at 0 K is $S_0 = 0$. As a result, the integration yields:
\begin{equation}
S=\int dS=\int\frac{dH}{T}=\int_{0}^{T}\frac{C_P(T)}{T}dT.
\label{Entropyformula}
\end{equation}
The entropy we determined for solid and liquid states are shown in FIG.~\ref{Entropy}. According to the results in FIG.~\ref{Cboth}, the specific heat of the liquid phase is consistently higher than that of the solid phase in the temperature range before the melting point. Entropy is derived from the integral of the specific heat capacity, so within each temperature range, the entropy of the liquid phase is always greater than that of the solid phase. This observation aligns with the greater disorder in the liquid's structure, confirming the higher entropy of the liquid phase.

Under the ensemble with a constant pressure and conservation of particle number, the Gibbs energy can be determined as:
\begin{equation}
G(T)=U(T)-T\int_{0}^{T}\frac{C_P(T)}{T}dT.
\label{G}
\end{equation}
And the Gibbs free energy depended on temperature of solid and liquid are shown in Figure~\ref{FreeEnergy}. The intersection point between the Gibbs free energy curves of liquid and solid state Ga at around 220 K, which indicates that the solid-liquid transition temperature of Ga is approximately 220 K from our thermodynamic results. When the temperature is above 220 K, the Gibbs free energy of liquid Ga is lower than that of solid state, it means the liquid state Ga can stable exist above this temperature point of 300 K at room temperature. This discrepancy arises because in Eq.~\eqref{Entropyformula}, we are unable to precisely determine the initial entropy, leading to a shift in the intersection point of the two curves. As shown in red dash line in FIG.~\ref{FreeEnergy}, we take away a $T\Delta S_0$ to make this intersection to 300 K. However, here $\Delta S_0$ is negative, which contradicts the intuition that the absolute entropy of a liquid should be greater than that of a solid. If a more accurate method can be employed to determine the initial entropy, it would allow for a more precise determination of the phase transition point.

\section{conclusions}
In summary, we develop a DP model that accurately describes the dynamic properties of Ga, including essential factors such as the melting point and mean square displacement. This successful representation confirms the precision of our DP model in depicting both solid and liquid Ga phases. The DP+QTB method is able to introduce nuclear quantum effects with low computation cost, allowing for accurate thermodynamic property calculations for both solid and liquid phases Ga. By accounting for nuclear quantum effects, we calculated the thermodynamic properties of solid and liquid Ga, providing accurate estimations of interval energy and specific heat. These calculations allowed for the computation of the Gibbs free energy of the solid and liquid phase, offering an energy-based perspective to describe the phase stability of liquids. Through the case of liquid Ga, we demonstrated the effectiveness of the DP+QTB method, showcasing its potential to offer valuable insights into the physical characteristics of disordered systems such as liquids and amorphous structures.

\begin{acknowledgments}
This work was supported by the National Key R$\&$D Program of China (Grants No. 2021YFA0718900 and No. 2022YFA1403000), the Key Research Program of Frontier Sciences of CAS (Grant No. ZDBS-LY-SLH008), the National Nature Science Foundation of China (Grants No. 11974365, No. 12204496, No. 5193101 and No. 52127803), the K.C. Wong Education Foundation (GJTD-2020-11), the “Pioneer” and “Leading Goose” R$\&$D Program of Zhejiang (No. 2022C01032), and the Science Center of the National Science Foundation of China (52088101).  
\end{acknowledgments}

\providecommand{\noopsort}[1]{}\providecommand{\singleletter}[1]{#1}%

\end{document}


\preprint{APS/123-QED}

\title{Supplementary Materials of\\Atomistic simulations of thermodynamic properties of liquid gallium from first principles}

\author{Hongyu Wu$^{1,2}$}
\author{Wenliang Shi$^{1,2}$}
\author{Ri He$^{1,2}$}
\author{Guoyong Shi$^{3}$}
\author{Chunxiao Zhang$^{4}$}
\author{Zhicheng Zhong$^{1,2}$}
\thanks{zhong@nimte.ac.cn}
\author{Run-Wei Li$^{1,2}$}

\affiliation{\\$^1$CAS Key Laboratory of Magnetic Materials and Devices $\&$ Zhejiang Province Key Laboratory of Magnetic Materials and Application
Technology, Ningbo Institute of Materials Technology and Engineering, Chinese Academy of Sciences, Ningbo 315201, China}

\affiliation{$^2$College of Materials Science and Opto-Electronic Technology, University of Chinese Academy of Sciences, Beijing 100049, China}

\affiliation{$^3$Department of Physics, Yantai University, Yantai, 264005, China}

\affiliation{$^4$School of Physics and Optoelectronic Engineering, Shandong University of Technology, Zibo, Shandong 255100, China}



\maketitle

\section{generation and validation of the deep potential}
\begin{figure}[htbp]
\centering
\includegraphics[width=0.55\textwidth]{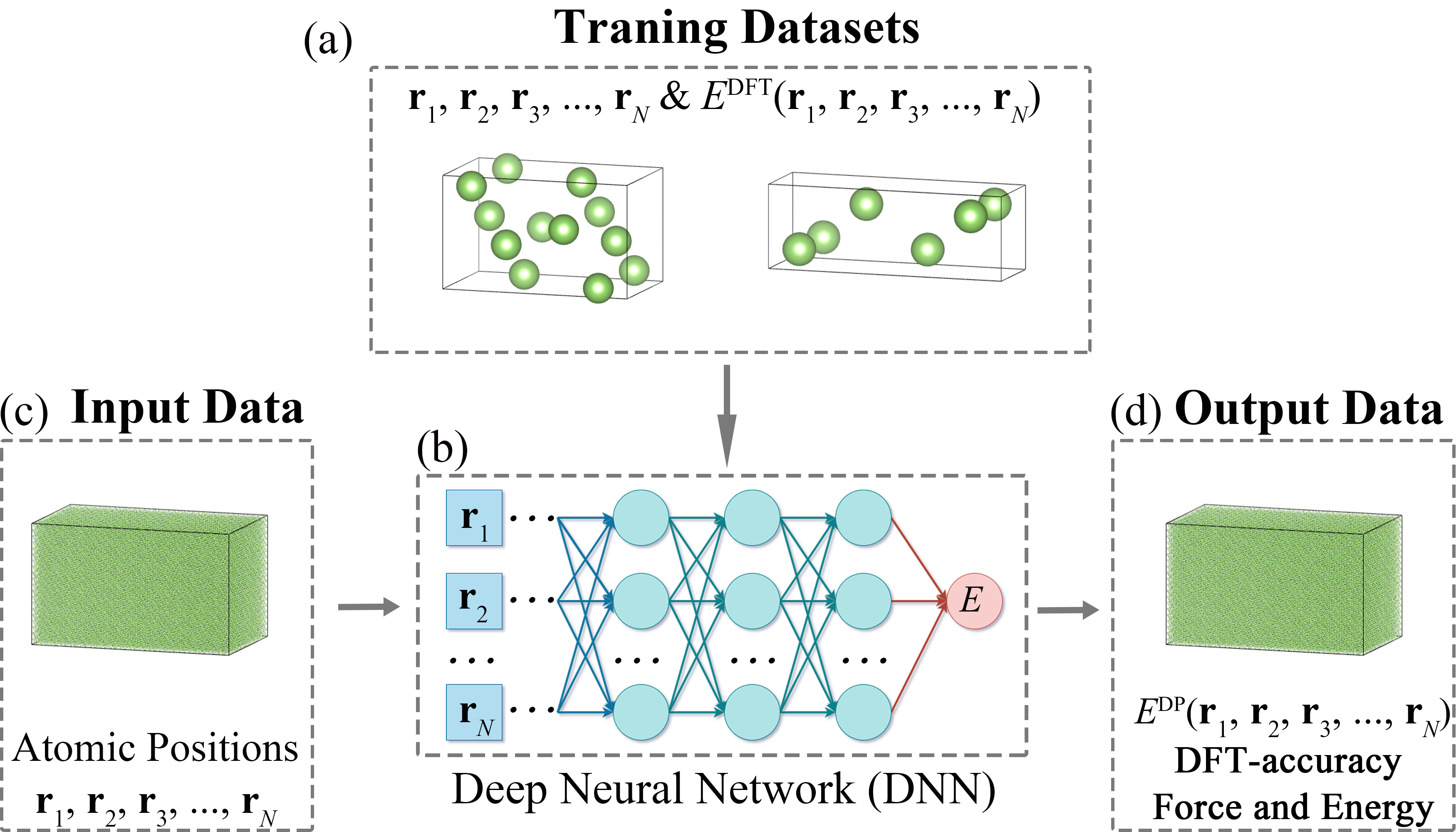}
\caption{(a) A large dataset of DFT energies and forces of a wide range of atomic configurations. (b) The deep neural network trained by Deep Potential. (c) and (d) The well-trained deep neural network can be applied to predict the energy and force of large supercell in DFT-level.}
\label{dp}
\end{figure}

Here we give some detail information about Deep Potential (DP)\cite{Zhang2018,Zhang2018a}. In the DP model, the total energy $E$ of a configuration is assumed to be the sum of “atomic energy” $E_i$ of each atom $i$. $E_i$ is constructed in two steps. First for each atom, their coordination and local environment within a cutoff radius $R_c$ will be remapped to a descriptor $D_i$ through an embedding network. The meaning of the descriptor is to preserve the symmetry of the system such as translation and rotation symmetries. In this work, the cutoff radius $R_c$ of descriptor is set to 6 $\AA$, the maximum number of atoms within the $R_c$ is set to 64. Next, the descriptor serve as input of a deep neural network called fitting network, which returns $E_i$ as output. (Schemmatic diagram is FIG.~\ref{dp}) The sizes of the embedding and fitting networks are (25, 50, 100) and (240, 240, 240), respectively. The parameters in fitting network are optimized by loss function, it is defined as
\begin{equation}
L(p_\epsilon,p_f,p_\xi)=p_\epsilon\Delta\epsilon^2+\frac{p_f}{3N}\sum_i \mid\Delta \mathbf{F}_i\mid^2 +\frac{p_\xi}{9} \mid\mid\Delta\xi\mid\mid^2,
\end{equation}
where all $\Delta$ denotes the difference between the DP predictions and the training datasets, $N$ is the number of atoms, $\epsilon$ is the energy per atom, $F_i$ is the force of atom $i$, $\xi$ is the virial tensor, $p_\epsilon$, $p_f$, $p_\xi$ is the weight coefficients for the three physical quantities, respectively. 

To generate a training dataset that covers a wide range of PES sufficiently, the DP generator (DP-GEN) is adopted\cite{Zhang2020}. Combining DP, Molecular Dynamics (MD) and Density Functional Theory (DFT) methods, DP-GEN can generate a large number of new configurations and corresponding DFT results based on the initial configuration in each round of iterations. The initial configuration is started with DFT relaxed ground-state structures of $\alpha$, $\beta$, II phase of Ga, as these three phases are the three fundamental structural phases observed experimentally for Ga. During the iteration, MD simulations adopt the isobaric-isothermal ensemble with temperature set from 50 to 600 K, pressure set from 0.001 to 50 kbar, and time with 1 to 8 picoseconds, the candidate configurations for final training dataset will be drawn in a specific way from these MD simulation results (For more details of the DP-GEN method, please refer to the Ref.~\cite{Zhang2020}). At the final, 4227 candidate configurations were selected for training dataset. The initial training dataset is obtained by performing a 20-step ab initio MD simulation for randomly perturbed structures at 50 K. After labeling candidate configurations, self-consistent DFT should be performed. All DFT calculations were performed using a plane-wave basis set with a cutoff energy of 500 eV, and the electron exchange-correlation potential was described using the generalized gradient approximation and PBEsol scheme\cite{Perdew1997}. The Brillouin zone was sampled with k-point grids of smallest allowed spacing between k points in 0.1 $\AA^{-1}$ for every configurations.

\section{Self diffusion coeffcient and melting point}

\begin{figure}[htbp]
\centering
\includegraphics[width=0.6\columnwidth]{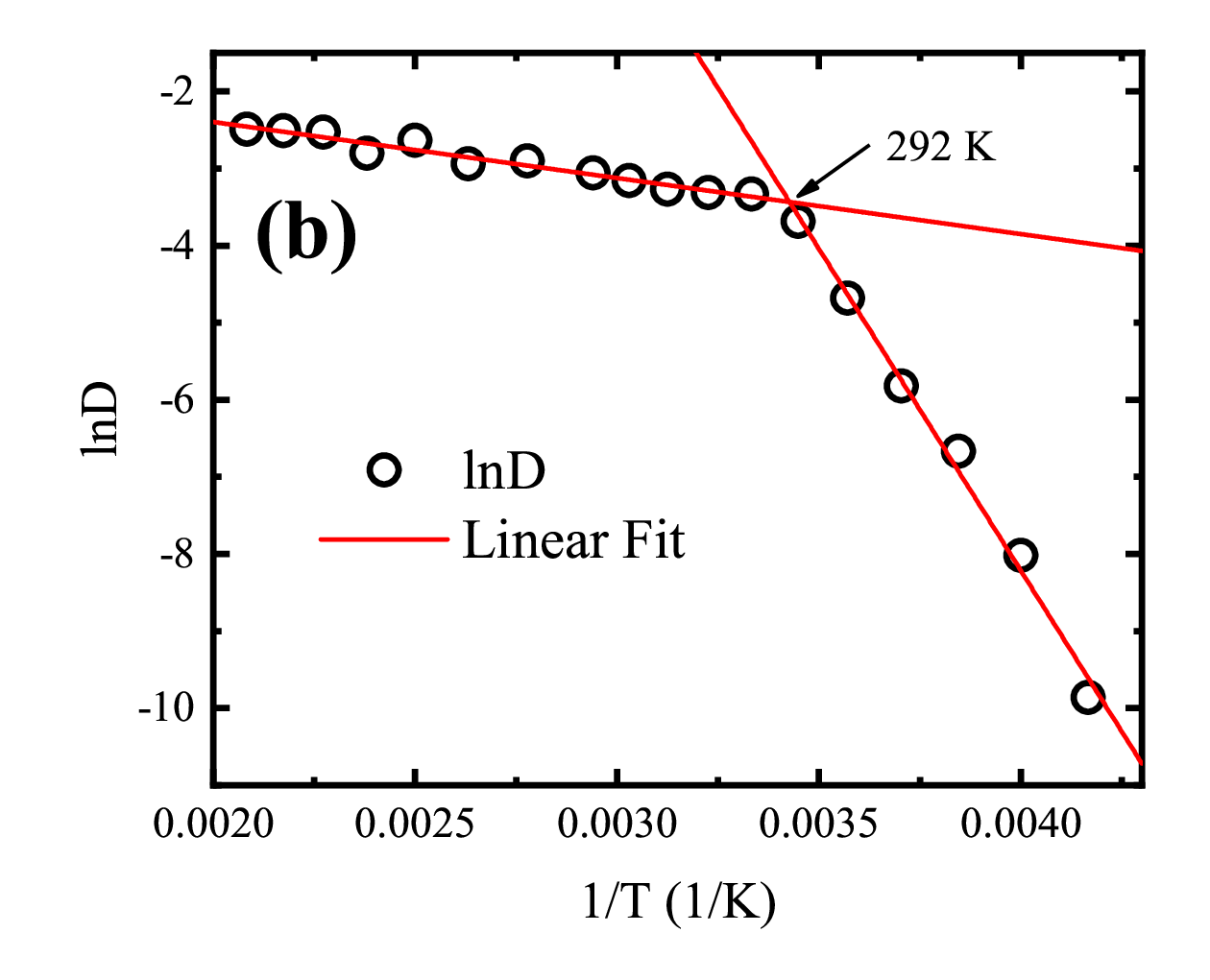}
\caption{The logarithm of the mean square displacement of $\alpha$-Ga, red lines are the linear fitting of two different temperatrue range. The intersection of the two red lines represents the solid liquid phase transition point}
\label{lnD}
\end{figure}

Given the high fluidity of liquid gallium, the collective thermal motion of gallium atoms can be studied by examining the diffusion properties resembling Brownian motion. To investigate the diffusion phenomenon observed in classical Brownian particles, an empirical Arrhenius formula is often employed to describe the diffusion coefficient:
\begin{equation}
D=D_0~exp\left(-\frac{Q}{k_BT}\right).
\end{equation}
In this equation, $D_0$ represents a constant, $k_B$ is the Boltzmann constant, $T$ denotes temperature, and $Q$ refers to the activation energy required for a particle to overcome the diffusion barrier. It is common to rewrite the equation in logarithmic form as follows:
\begin{equation}
\ln D=-\frac{Q}{k_BT}+\ln D_0
\end{equation}

It is generally believed that the values of $D_0$ and $Q$ are independent of temperature but are instead determined by the diffusion mechanism and material properties. As a result, a linear relationship exists between $lnD$ and $1/T$. Clearly, when atoms diffuse through two different mechanisms at high and low temperatures, there will be two distinct line segments with different slopes in the $lnD$-$1/T$ plot due to the different diffusion activation energies. For gallium, these two diffusion mechanisms correspond to the liquid state at high temperatures and the solid state at low temperatures. The turning point in the plot corresponds to the temperature at which the phase transition occurs. As shown in FIG.~\ref{lnD}, the $lnD$-$1/T$ plot exhibits two linear segments, representing the slight thermal motion of atoms around their equilibrium positions in the solid phase and the vigorous diffusion of atoms after melting. By performing linear fitting on the two segments of data, the intersection point of the two lines is determined to be at 292K, which is remarkably close to the melting point of gallium under atmospheric pressure.

\section{Macroscopic and microscopic structures of gallium}

\begin{figure}[htbp]
\centering
\includegraphics[width=0.6\columnwidth]{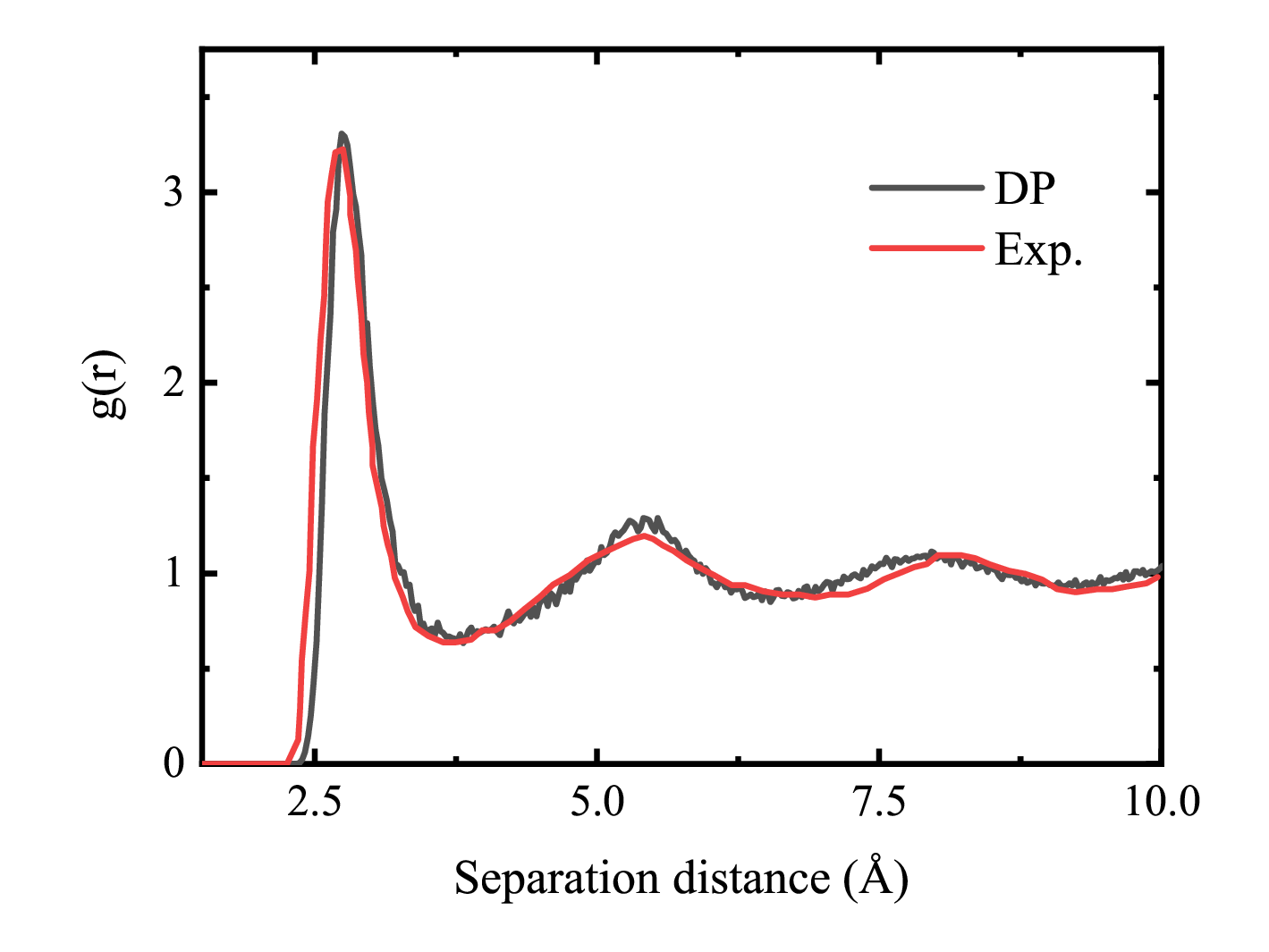}
\caption{Radial distribution function of Ga at 300 K and
1 atm, which are predicted by DP model and obtained from
experience, respectively.}
\label{RDF}
\end{figure}

For the liquid structure obtained from our DP model, we also compared it with experimental results. The radial distribution function (RDF) obtained from the DP model is shown in FIG.~\ref{RDF}. The overall trend of the RDF obtained from the DP model aligns with the experimental results. However, there is a slight shift in the position of the first peak compared to the experimental value (2.712 Å vs. 2.738 Å). This discrepancy can be attributed to the use of the Perdew-Burke-Ernzerhof (PBE) functional during the training of the DP model, which tends to overestimate the lattice constant compared to experimental values.

\begin{figure}[htbp]
\centering
\includegraphics[width=0.6\columnwidth]{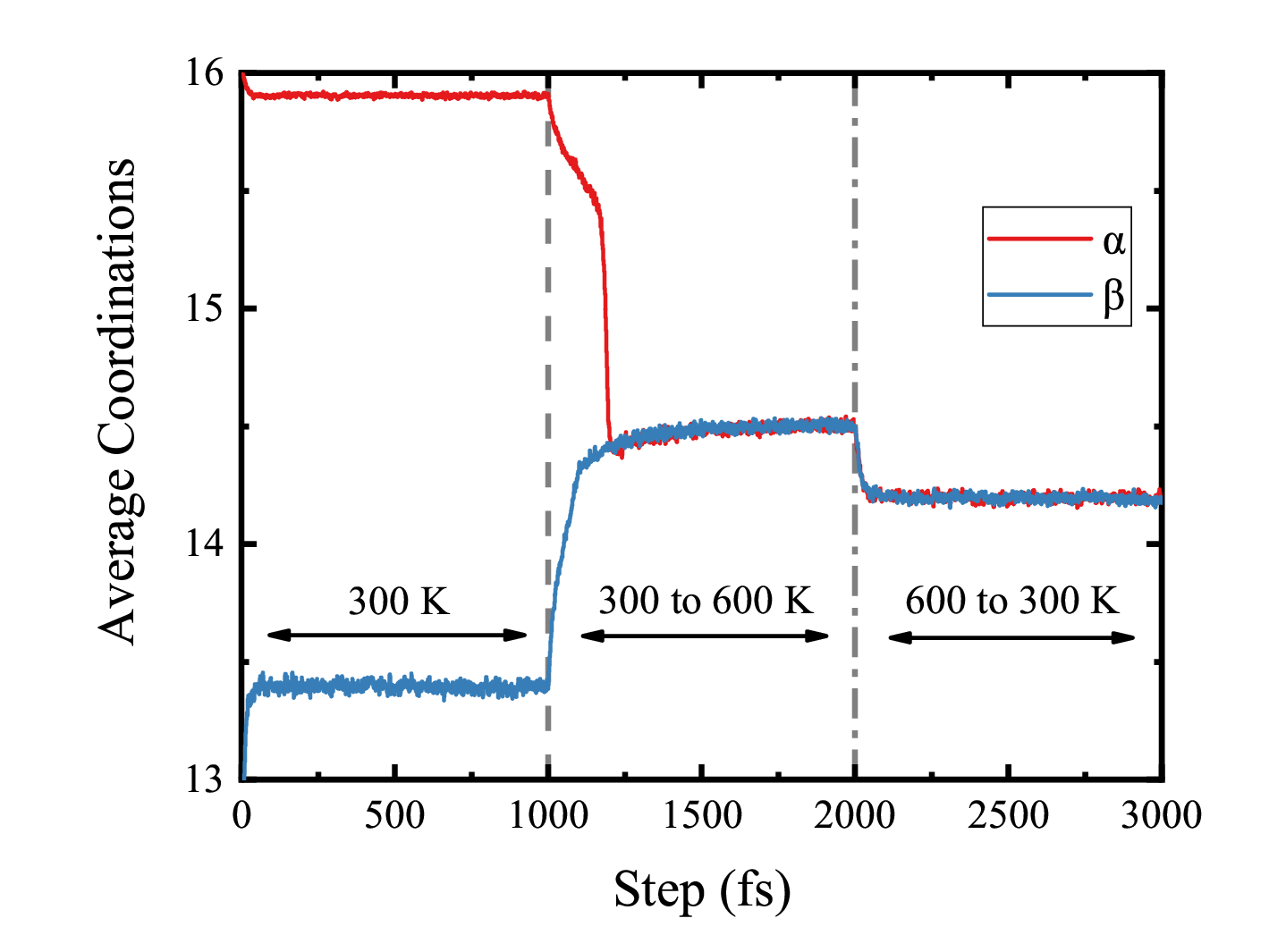}
\caption{Coordination of $\alpha$ and $\beta$ gallium, the simulation are
both started at low temperature. Dashed line represents heating
and dashed dot line represent cooling.}
\label{Coord}
\end{figure}

The results of the diffusion coefficient and RDF indicate that, overall, the liquid gallium transition described by the DP model closely resembles the liquid structure and actual behavior of gallium. The above results demonstrate that our DP model exhibits excellent accuracy, as it can capture the complex interactions in gallium, describe its disordered liquid structure, and account for the effects induced by temperature and pressure variations. Furthermore, we delve into the microscopic atomic-scale to describe certain phenomena. We selected gallium in the $\alpha$ and $\beta$ phases as the initial structures for performing molecular dynamics simulations. For each phase, the simulations involved heating the system above the melting point and subsequently cooling it to the supercooled liquid (SCL) state. We then averaged the coordination numbers of each atom and presented the results in Figure~\ref{Coord}. Initially, due to the distinct crystalline environments of the $\alpha$ and $\beta$ phases, there is a noticeable difference in their coordination numbers. However, as the temperature is raised to the melting point, the coordination numbers gradually converge towards a similar value for both phases. Upon cooling to the supercooled liquid state, the coordination numbers of both phases remain consistent. This indicates that the distinctive structural characteristics of the different gallium phases are not preserved after melting, resulting in a convergent state.

\begin{figure}[htbp]
\centering
\includegraphics[width=0.6\columnwidth]{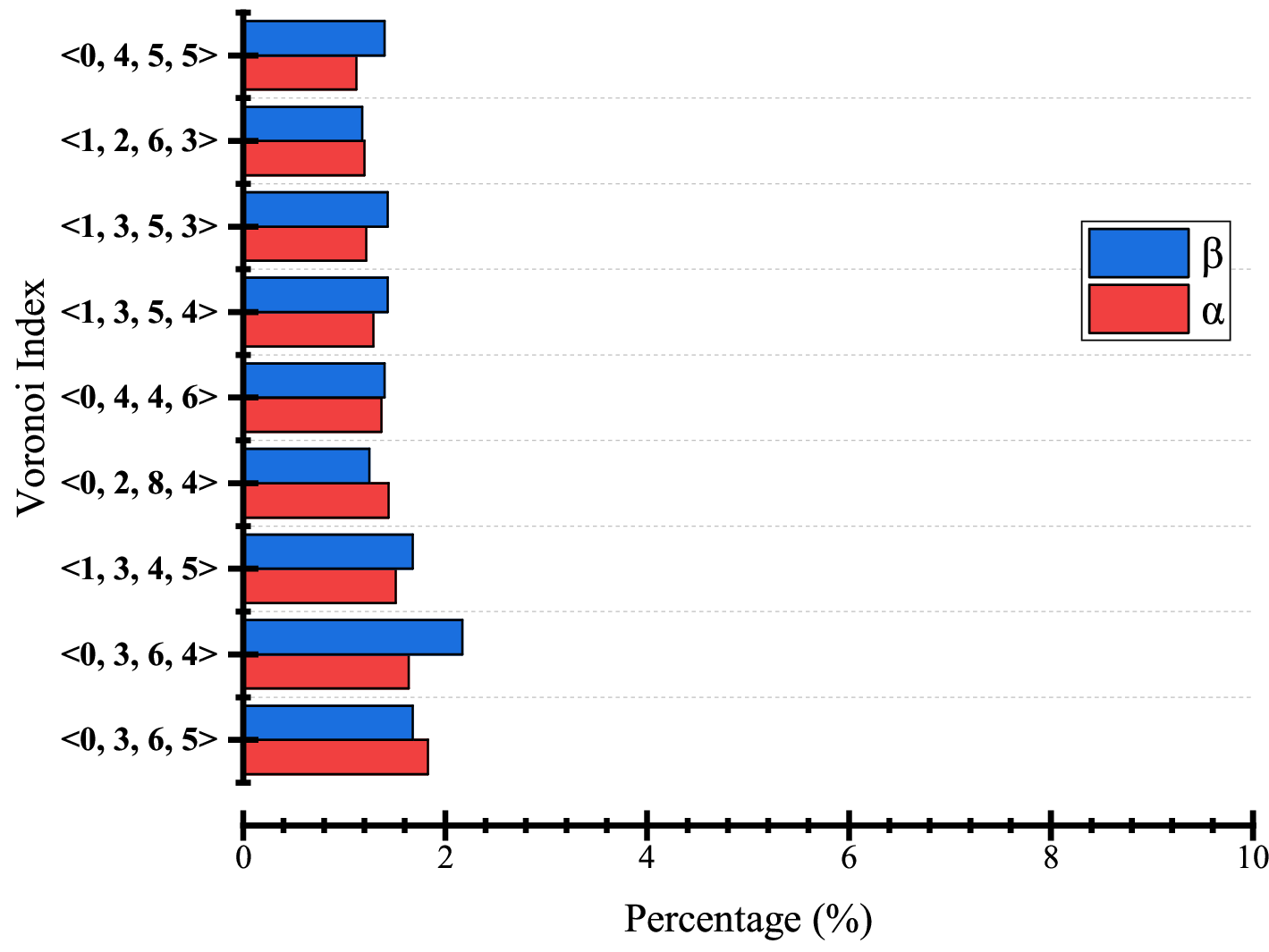}
\caption{Major Voronoi index for both liquid originate from $\alpha$ and $\beta$ phases.}
\label{Voronoi}
\end{figure}

We further analyzed the microstructural features of these phases using Voronoi analysis, which partitions space into polyhedral cells based on the perpendicular bisector planes formed by each atom and its nearest neighbors. Our analysis revealed that no dominant polyhedral type was observed in either supercooled liquid phase. Instead, a variety of polyhedra were uniformly present. This suggests that liquid gallium exhibits a highly disordered and uniform structure without any short-range ordering and lacks multiple types of liquid structures. For the specific table containing the polyhedral indices, please refer to the FIG.~\ref{Voronoi}.

\begin{figure}[htbp]
\centering
\includegraphics[width=0.6\columnwidth]{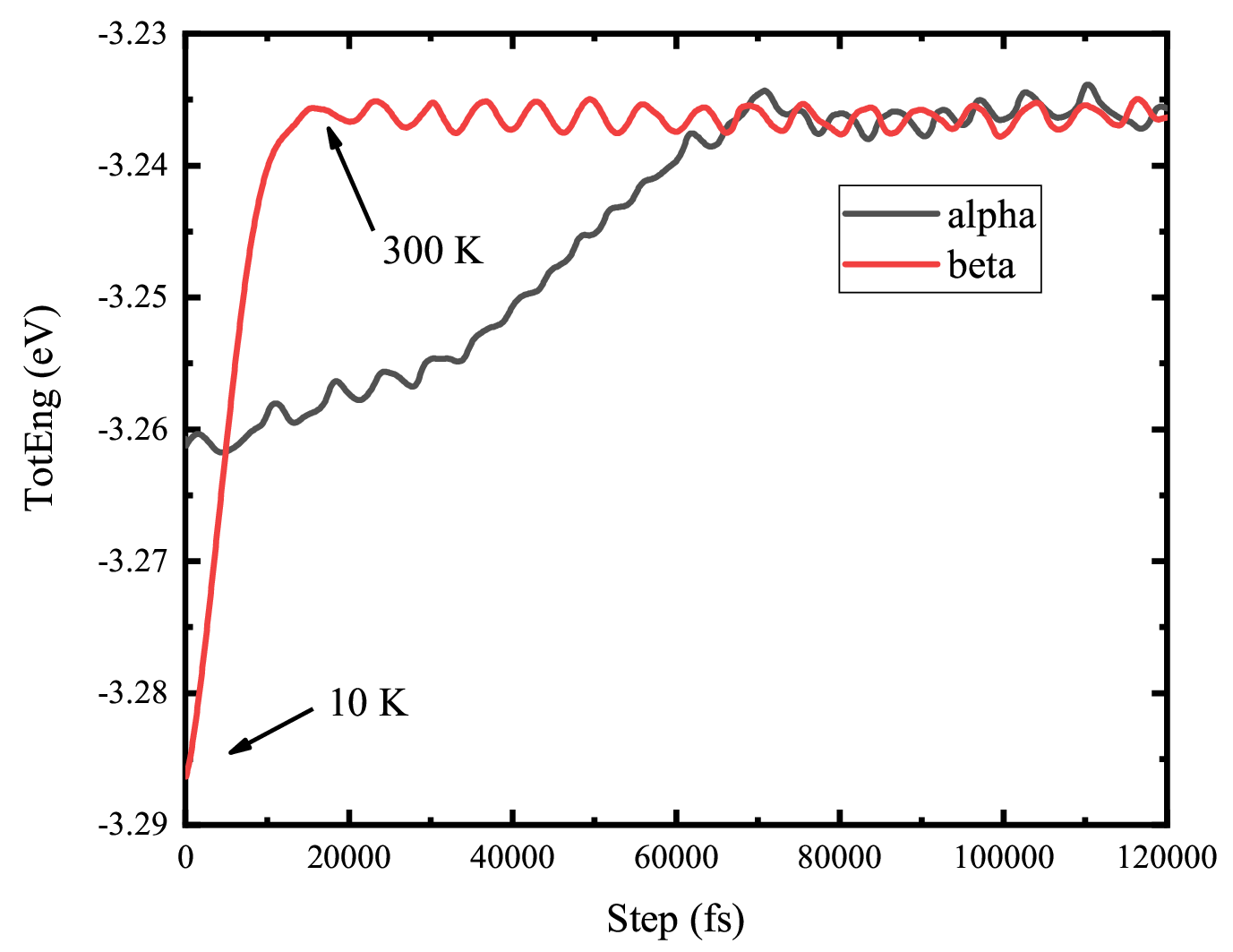}
\caption{Energy of $\alpha$ and $\beta$ Ga during simulation, the initial temperature is set to 10 K and final equilibrium temperature is 300 K.}
\label{abenergy}
\end{figure}  

Furthermore, the same conclusion can be corroborated from an energy perspective. We constructed structures for both $\alpha$ and $\beta$ phases with melting nuclei and simulated them at a constant temperature of 300K. As shown in the Fig.~\ref{abenergy}, the results indicate that the liquids obtained from melting both structures exhibit consistent energy behavior, implying that they possess similar microstructural characteristics.

\begin{figure}[htbp]
\centering
\includegraphics[width=0.6\columnwidth]{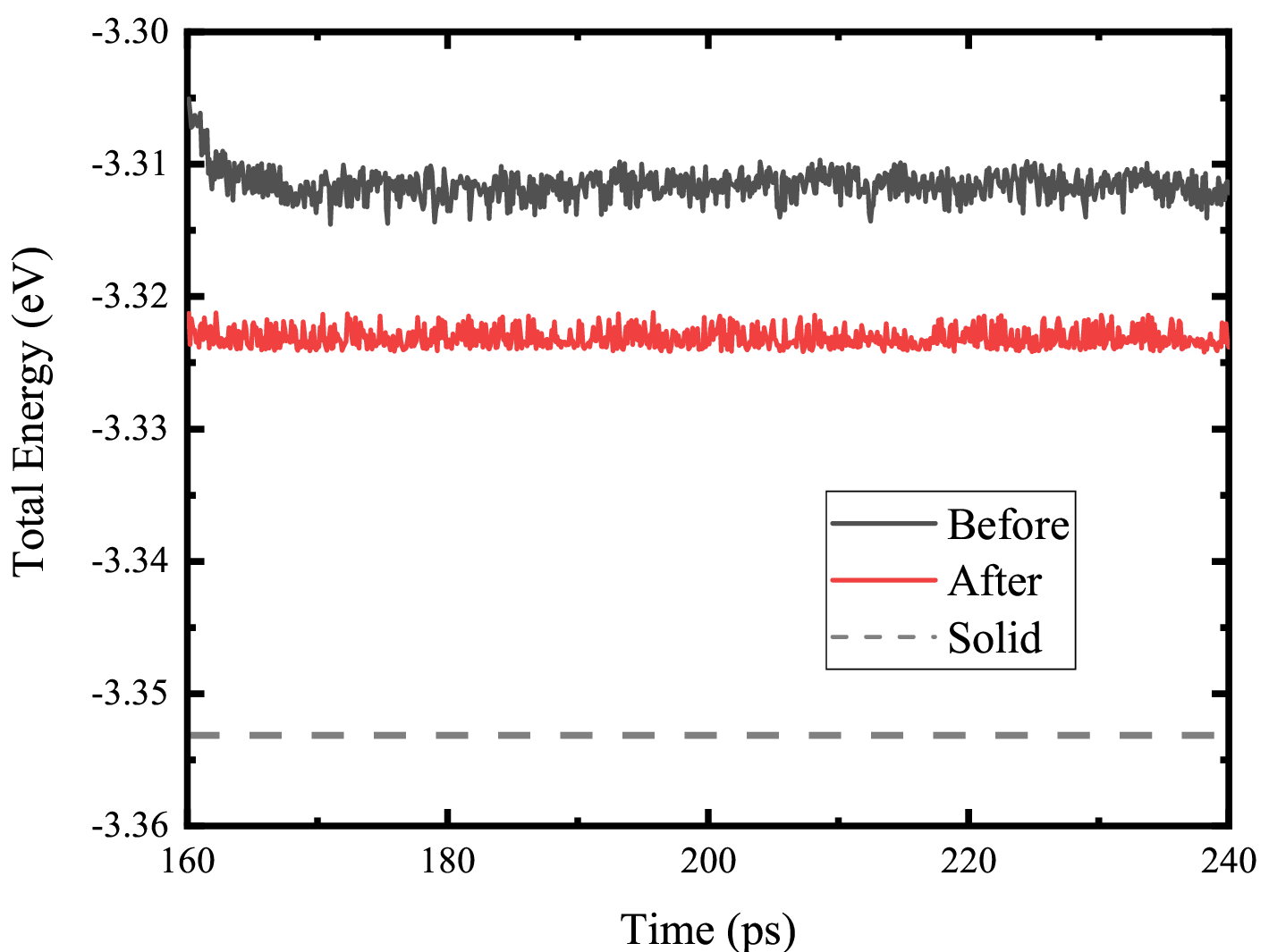}
\caption{Energy of liquid (before and after relaxation) and solid states.}
\label{RelaxEnergy}
\end{figure}

Finally, we compare the energy between liquid and solid states, the energy of relaxed liquid states are higher than solid states about 30 meV/atom, as shown in FIG.\ref{RelaxEnergy}.

\section{Vibration modes}

\begin{figure}[htbp]
\centering
\includegraphics[width=0.6\columnwidth]{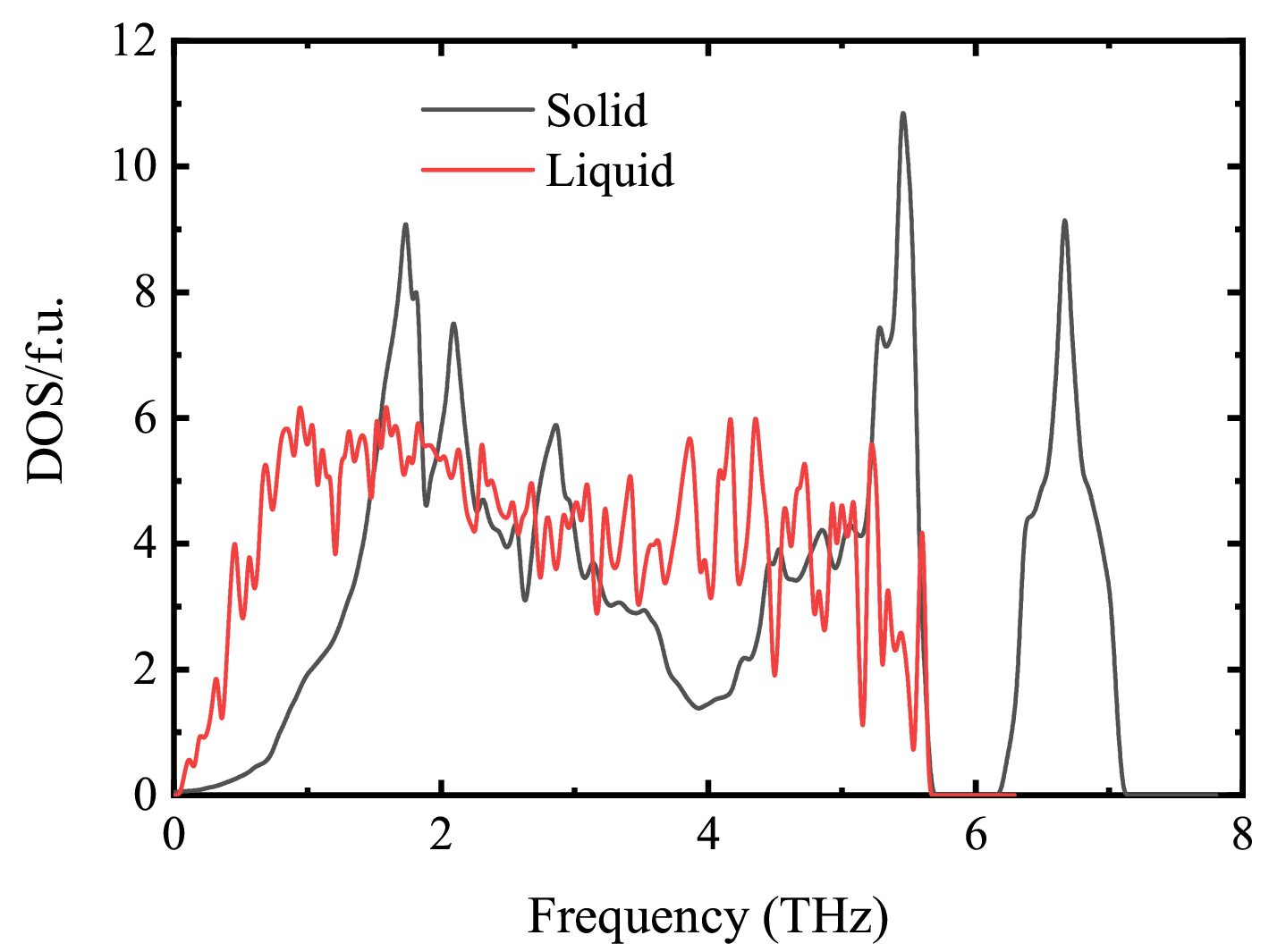}
\caption{Comparison of phonon density of states of solid and liquid gallium}
\label{Phonon}
\end{figure}

We employed the DP+QTB method in combination with density functional perturbation theory method to calculate the atomic vibration modes of solid and liquid Ga, as shown in FIG.\ref{Phonon}. The phonon density of states (PDOS) for the solid phase Ga is based on $\alpha$ phase, and liquid phase is based on a supercell contain 216 atoms. Compared to the solid Ga, the PDOS of liquid Ga show a higher excitation of low-frequency modes, while their high-frequency phonons disappear simultaneously. This indicates that when the structure of gallium becomes disordered, the high-frequency vibrations that were originally near the equilibrium positions decrease, while the atomic diffusion and mobility increase. Furthermore, the integrated density of states in the liquid phase is also smaller than that in the solid phase, which indicates that the $T_{zero}$ in the liquid phase is smaller than in the solid phase, and this might be one of the origins of difference of thermodynamics between solid and liquid phase.


\providecommand{\noopsort}[1]{}\providecommand{\singleletter}[1]{#1}%